\documentclass{aastex631}

\usepackage{bm}
\usepackage{amsthm}
\usepackage{amsfonts}
\usepackage{anysize}
\usepackage{amsmath}
\usepackage{mathtools}
\usepackage[english]{babel}
\usepackage{graphicx}
\usepackage{amssymb}

\newcommand{\be}{\begin{equation}}
\newcommand{\ee}{\end{equation}}
\newcommand{\ba}{\begin{eqnarray}}
\newcommand{\ea}{\end{eqnarray}}
\newcommand{\non}{\nonumber}
\newcommand{\n}[1]{\label{#1}}

\newcommand{\is}{{\text{\tiny ISCO}}}
\newcommand{\ce}{{\cal{E}}}
\newcommand{\cl}{{\cal{L}}}

\newcommand{\nin}{\noindent}

\begin{document}

\title{Charged Particles Circular Orbits around Weakly Charged and Magnetized Kerr Black Holes}

\email{ama3@ualberta.ca, amz@kfupm.edu.sa}
\author[0000-0002-5142-6904]{A. M. Al Zahrani}
\affiliation{Physics Department, King Fahd University of Petroleum and Minerals \\ Dhahran 31261, Saudi Arabia}

\begin{abstract}

We study the circular orbits of charged particles around a weakly charged Kerr black hole immersed in a weak, axisymmetric magnetic field. Firstly, we review the circular orbits of neutral particles. We then review the circular orbits of charged particles around a weakly charged Kerr black hole and weakly magnetized Kerr black hole. The case of a weakly magnetized and charged black hole is investigated thereafter. We investigate, in particular, the effect of the electromagnetic forces on the charged particles' innermost stable circular orbits. We examine the conditions for the existence of negative-energy stable circular orbits and the possibility of the emergence of a gap or double orbit in thin accretion disks. Some of the interesting astrophysical consequences of our findings are discussed as well.

\end{abstract}


\section{Introduction}

Black holes are among the most fascinating objects to explore. We study them via their interaction with other stars or through the matter surrounding them; their accretion disks. The physical properties of a black hole have direct impact on its accretion disk. Therefore, studying the accretion disk can tell us a lot about the black hole it engulfs and about the disk itself. 

The no-hair-theorem states that a stationary black hole is characterized by only three parameters, mass, spin angular momentum and charge~\citep{MTW}. However, it is widely adopted that astrophysical black holes are neutral. This is because the selective accretion of the ambient matter would shortly neutralize any excess charge. Therefore, we can hardly find astrophysical charged black holes investigated in the literature, except for academic purposes. 

Although the assertion about black hole charge is well-founded, weakly charged black holes are not unlikely to exist. There are several profound reasons that back their existence. The mass difference between electrons and protons may render a black hole positively charged (see \citep{Zaj1,Zaj2,Zaj3} and the references within). Furthermore, a rotating black hole immersed into a homogeneous magnetic field acquires a non-vanishing electrical field. This can result in the selective accretion of the ambient free charged particles, until the black hole's charge neutralizes the magnetically-induced electric field \citep{Carter, Wald}. Another charging mechanism is due to the difference in the accretion rates of electrons and protons of the surrounding plasma in the presence of radiation from the accreting matter.

Nowadays, it is confirmed that astrophysical black holes are magnetized. The magnetic fields are mainly generated by the plasma in the accretion disk as discussed in \cite{RoVi,Pun}. There are several observational measurements of black hole magnetic fields \citep{Ruth,sil1,sil2,sil3,sil4}. 

One of the most informational parameters of a black hole accretion disk is the innermost stable circular orbit (ISCO). It is vital for measuring the spin angular momentum of the black hole \citep{Bren}. Moreover, it is a major determinant of the structure of the accretion disk \citep{AF}. The ISCO coincides with the inner edge of an accretion disk when its luminosity is low compared to the Eddington luminosity \citep{Ab}. The ISCO has therefore a direct impact on the appearance of a black hole's shadow.

The charged particles' ISCOs around a Schwarzschild and Kerr black holes immersed in a weak, axially symmetric magnetic field were extensively investigated \citep{GP,AG,AO,Z}. In all cases, the effect of the magnetic field is to bring the ISCO closer to the black hole. The dynamics of charged particles was investigated in \cite{PQR,PQR2,RFSJ,SG} for the Reissner-Nordström spacetime and in \cite{SG} for the Kerr-Newman spacetime. The charged particles' ISCOs around a weakly charged Schwarzschild black hole were addressed in \cite{Zaj1,Zaj2,Zaj3,Z2}. When the charge is weak, the effect of the Coulomb force is to push the ISCO farther away whether the force is attractive or repulsive. When the force is repulsive, there is a maximum value of the force after which the ISCO ceases to exist. 

The combined affects of the magnetic field and black hole's charge on the ISCO of a Schwartzschild black hole were addressed in \cite{HH} for a few special cases. We investigated the problem further in \cite{Z2}. We found that negative energy ISCOs become possible. More importantly, we found that two bands of circular orbits separated by gap of no stable circular orbits can exist. 

This paper is a generalization of our work in \cite{Z2}. Namely, we study the ISCOs of charged particles orbiting a weakly charged Kerr black hole, immersed in a weak, axisymmetric magnetic field. The black hole's charge and magnetic field are weak in the sense that their back-reactions on the spacetime are insignificant\footnote{In what follows, it should be inferred that the black hole's charge and magnetic field are assumed to be weak}. We start by inspecting the effect of each of the black hole's charge and magnetic field separately, and then study their combined effect. The possibility of the existence of a region with no stable circular orbits or double stable circular orbits is demonstrated. The existence of negative-energy stable circular orbits is investigated in some detail. The paper is organized as follows: Sec.~\ref{s2} is a review of the circular orbits and the ISCO of a neutral particle. In Sec.~\ref{s3}, we  review the Wald's solution of Maxwell's equations in Ricci flat spacetimes and write the radial equation of motion for a charged test particle. In Secs.~\ref{s4} and \ref{s5} we discuss the ISCO of a charged particle near a charged Kerr black hole and magnetized black hole, respectively. In Sec.~\ref{s6}, we investigate in detail the charged particle's ISCO near a charged Kerr black hole immersed in a magnetic field. General discussion and conclusion is given in Sec.~\ref{sum}. We use the sign conventions adopted in \cite{MTW} and geometrized units where $c$, $G$ and $k$ (the Coulomb constant) are unity.

\section{Circular Orbits Of A Neutral Particle Around A Kerr Black Hole} \label{s2}

The spacetime geometry around a spinning black hole of mass $M$ and spin angular momentum $J = aM$ is described by the Kerr metric. In the Boyer-Lindquist coordinates, it reads~\citep{MTW}
\begin{eqnarray}
ds^2=-\frac{\Delta}{\Sigma}\left(dt-a\sin^2{\theta}\;d\phi\right)^2 d\theta^2+\frac{\sin^2{\theta}}{\Sigma}\left[adt-(r^2+a^2)\;d\phi\right]^2
\end{eqnarray}
\nin where $\Sigma = r^2+a^2\cos^2\theta$, $\Delta = r^2+a^2-2Mr$, and $a$, with $-M\leq a\leq M$, is the spin parameter. The Kerr metric admits two Killing vectors and a Killing tensor. The two Killing vectors are the purely temporal and azimuthal Killing vectors, since the metric is temporally and azimuthally symmetric. They respectively read
\begin{equation}
\xi^{\mu}_{(t)}=\delta_t^{\mu}, \:\:\: \xi^{\mu}_{(\phi)}=\delta_{\phi}^{\mu}.
\end{equation}

\nin Let a test particle of mass $m$ be moving with four-velocity $u^{\mu}$ in the Kerr background. There are two constants of the particle's motion associated with the two Killing isometries:
\begin{eqnarray}
\ce &=& -p_{\mu}\xi^{\mu}_{(t)}/m = \left(1-\frac{2\text{Mr}}{\Sigma}\right)\dot{t}+\frac{2aMr\sin^2\theta}{\Sigma}\dot{\phi}, \label{en0} \\
\cl &=& p_{\mu}\xi^{\mu}_{(\phi)}/m = \frac{\Gamma\sin^2{\theta}}{\Sigma}\dot{\phi}-\frac{2aMr\sin^2\theta}{\Sigma}\dot{t},\label{am0}
\end{eqnarray}
\nin where $\Gamma=(r^2+a^2)^2-\Delta a^2 \sin^2\theta$. Here $p^{\mu}=mu^{\mu}$ is the particle's four-momentum. The two constants of motion $\ce$ and $\cl$ are the specific energy and specific azimuthal angular momentum, respectively. Using them along with the normalization $u_{\mu}u^{\mu}=-1$, we reduce the radial equation of motion in the equatorial submanifold ($\theta=\pi/2$, $\dot{\theta}=0$) to quadrature:
\begin{equation}
r^4\dot{r}^2=\left[(r^2+a^2)\ce-a\cl\right]^2-\Delta r^2 \label{rdot0}.
\end{equation}
\nin The overdot denotes differentiation with respect to the particle's proper time. The radial motion is  invariant under the transformations
\begin{equation}\label{sym}
\phi\rightarrow-\phi, \hspace{3mm} \dot{\phi}\rightarrow-\dot{\phi}, \hspace{3mm} \cl\rightarrow-\cl \hspace{3mm} a\rightarrow -a.
\end{equation}

\begin{figure}[h!]
  \centering
  \includegraphics[width=0.45\textwidth]{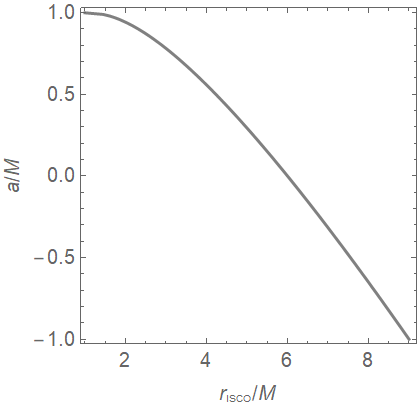}
  \caption{The dependence of the innermost stable circular orbit radius $r_\is$ for a neutral particle on the spin parameter $a$.}\label{f:rvsa}
\end{figure}

\nin Therefore, there are {\it two} modes of radial motion, depending on whether the black hole's spin and the particle's azimuthal angular momentum are parallel ($a\cl>0$) or anti-parallel ($a\cl<0$). Without loss of generality, we will keep $\cl$ positive while $a$ can take both signs.
\nin Let us define $R(r)$ to be the right hand side of Eq.~\ref{rdot0}:
\begin{equation}
R(r):=\left[(r^2+a^2)\ce-a\cl\right]^2-\Delta r^2.
\end{equation}
\nin $R(r)$ is positive semidefinite; it vanishes at the radial turning points only. Stable circular orbits exist where $R(r)$ and its first derivative $R'(r)$ vanish, or
\begin{eqnarray}
&&\left[(r^2+a^2)\ce-a\cl\right]^2-\Delta r^2=0,\\
&&r\left[(r^2+a^2)\ce-a\cl\right]\ce-r\Delta-(r-M)r^2=0.
\end{eqnarray}
\nin Solving these equations for $\ce$ and $\cl$ we obtain
\begin{eqnarray}
\ce_{\text o}&=&\frac{(r_{\text o}-M)r_{\text o}+\Delta_{\text o}}{2r_{\text o}\Delta_{\text o}^{1/2}},\label{ec}\\
\cl_{\text o}&=&\frac{r_{\text o}^2+a^2}{a}\ce_0-\frac{a\Delta_{\text o}^{1/2}}{r_{\text o}}.\label{lc}
\end{eqnarray}
\nin We have used $r_{\text o}$, $\ce_{\text o}$, $\cl_{\text o}$ and $\Delta_{\text o}$ to denote quantities corresponding to circular orbits. The value of $\ce_{\text o}$ is always positive with a value of $1$ far away from the black hole.
\nin A circular orbit is the ISCO when $R''(r_{\text o})$ vanishes, or
\begin{eqnarray}
(6r_{\text o}^2+a^2)(\ce_{\text o}^2-1)+6Mr_{\text o}-\cl_{\text o}^2=0.
\end{eqnarray}
\nin Plugging the $\ce_{\text o}$ and $\cl_{\text o}$ expressions above in this condition yields
\begin{equation}
r_\is(r_\is-6M)+8a\sqrt{Mr_\is}-3a^2=0.
\end{equation}
\nin The $\ce_{\text o}$ and $\cl_{\text o}$ expressions reduce for the ISCO to
\begin{eqnarray}
\ce_\is^2&=&1-\frac{2}{3}\frac{M}{r_\is}, \\
\cl_\is^2&=&\frac{2}{3}\frac{M}{r_\is}(3r^2_\is-a^2).
\end{eqnarray}
\nin Figure~\ref{f:rvsa} shows how $r_\is$ changes with $a$. The ISCO radius lies in the interval $[M,9M]$. When $r_\is=M$, $\ce_\is$ is minimum with a value of $1/\sqrt{3}$. Therefore, a particle ending in the ISCO can release an energy of up to $1-1/\sqrt{3}\approx 0.42$ of its rest energy.

\section{Charged and Magnetized Kerr Black Holes}\label{s3}

Let us now review Wald's solution of the Maxwell equations in a curved spacetime for weak electromagnetic fields introduced in \cite{Wald}. In a Ricci-flat spacetime a Killing vector $\xi^{\mu}$ obeys the equation
\begin{equation}
\xi^{\mu\:\:\:\: ;\nu}_{\:\:\;;\nu}=0.
\end{equation}

\noindent This is identical to the source-free Maxwell equations for a four-potential $A^{\mu}$ in the Lorentz gauge ($A^{\mu}_{\:\:\:;\mu}=0$),
\begin{equation}
A^{\mu\:\:\:\: ;\nu}_{\:\:\;;\nu}=0.
\end{equation}
Therefore, any linear combination of the Killing vectors the spacetime admits is automatically a solution to the Maxwell equations.

\nin In particular, the electromagnetic potential constructed of the temporal and azimuthal Killing vectors of the Kerr spacetime
\begin{equation}
A^{\mu}=\left(aB-\frac{Q}{2M}\right)\xi^{\mu}_{(t)}+\frac{B}{2}\xi^{\mu}_{(\phi)},
\end{equation}
\noindent describes the electromagnetic fields around a charged Kerr black hole immersed in an axisymmetric magnetic field \citep{Wald}. The black holes's charge $Q$ is given by 
\begin{equation}\label{empot}
Q=\frac{1}{4\pi}\int_{\sigma}F^{\mu\nu}d\sigma_{\mu\nu},
\end{equation}
\nin where $\sigma$ is a 2D surface surrounding the black hole and $F^{\mu\nu}$ is the electromagnetic field tensor (see Eq.~\ref{emt}). The magnetic field is axisymmetric with a strength of $B$ asymptotically \citep{Wald,AG,AO}. This is the potential that we will use in this paper. 

The dynamics of a charged particle of mass $m$ and charge $e$ in an electromagnetic field in a curved spacetime is governed by the equation
\begin{equation}\label{de}
m{u}^{\nu}\nabla_{\nu}u^{\mu}=eF^{\mu}_{\;\;\rho}u^{\rho}.
\end{equation}

\noindent The electromagnetic field tensor $F_{\;\;\nu}^{\mu}$ is given by
\begin{equation}\label{emt}
 F_{\mu\nu}=A_{\nu,\mu}-A_{\mu,\nu}.
\end{equation}

\nin In the frame of an observer with four-velocity $u^\mu_{\text{(obs)}}$, the electric and magnetic fields are, respectively
\begin{eqnarray}
E^{i}&=&F^{i\nu}u_{\nu\text{(obs)}}, \\
B^{\mu}&=&\frac{1}{2}\frac{\varepsilon^{\mu\nu\lambda\sigma}}{\sqrt{-g}}F_{\lambda\sigma}u_{\nu\text{(obs)}},
\end{eqnarray}
where $g=\mbox{det}(g_{\mu\nu})$, $\varepsilon_{0123}=+1$ and $i = 1,2,3$.
\nin For a static observer (which can exist outside the ergosphere only),
\begin{equation}
u^\mu_{\text{(obs)}}=(-g_{tt})^{1/2}\delta^{\mu}_t,
\end{equation}
\nin and the electric and magnetic fields in the equatorial plane are, respectively,
\begin{eqnarray}
E^i&=&\frac{Q-aBM}{r^2\sqrt{1-\frac{2M}{r}}}\delta_r^i, \\
B^i&=&\frac{a(Q-aBM)-Br^2(r-2M)}{r^4\sqrt{1-\frac{2M}{r}}}\delta_\theta^i.
\end{eqnarray}
\nin It is noteworthy that the electric field does not vanish when $Q=0$ as long as $aB\neq 0$. This is because the rotation of the black hole in the presence of the magnetic field induces an electric field. Typically, this induced electric field leads to selective accretion of the ambient charged particles until $Q=aBM$. 
\nin The generalized four-momentum of the particle is
\begin{equation}
P_{\mu}=mu_{\mu}+eA_{\mu}.
\end{equation}
\nin The Lie derivatives of $A_{\mu}$ with respect to $\xi_{(t)}^{\mu}$ and $\xi_{(\phi)}^{\mu}$ identically vanish:
\begin{eqnarray}
{\cal L}_{\xi^{\nu}_{(t)}}A^{\mu}&=&0, \\
{\cal L}_{\xi^{\nu}_{(\phi)}}A^{\mu}&=&0.
\end{eqnarray}
\nin The energy and azimuthal angular momentum of a charged particle are therefore constants of motion. Eqs.~\ref{en0} and~\ref{am0} are generalized to 
\begin{eqnarray}
{\cal E} &=& -P_{\mu}\xi^{\mu}_{(t)}/m = \frac{(q-4abM)r}{\Sigma}+\left(1-\frac{2Mr}{\Sigma}\right)\dot{t}+\frac{2aMr}{\Sigma}(b+\dot{\phi})\sin^2{\theta},  \label{en1}\\
{\cal L} &=& P_{\mu}\xi^{\mu}_{(\phi)}/m =\left[\frac{a(q-4abM)r}{\Sigma}-\frac{2aMr}{\Sigma}\dot{t}+\frac{\Gamma}{\Sigma}(b+\dot{\phi})\right]\sin^2{\theta}. \label{am1}
\end{eqnarray}
\nin where $q={eQ}/{m}$ and $b={eB}/{2m}$. We can straightforwardly obtain the charged particle's version of Eq.~\ref{rdot0} by combining Eqs.~\ref{en1} and~\ref{am1} with $u_{\mu}u^{\mu}=-1$. The radial equation of motion in the equatorial submanifold then reads
\begin{eqnarray}
r^3\dot{r}^2=\varepsilon^2\frac{\Gamma_0}{r}+4aM\varepsilon\lambda-r\left[\left(1-\frac{2M}{r}\right)(r^2+\lambda^2)+a^2\right], \label{rdot}
\end{eqnarray}
\nin where $\Gamma_0=\Gamma|_{\theta=\frac{\pi}{2}}=r^4+a^2r(r+2M)$ and
\begin{eqnarray}
\varepsilon &=& \frac{q-2abM}{r}-\ce, \\
\lambda &=& \frac{(q-4abM)a}{r}-b\frac{\Gamma_0}{r^2}+\cl.
\end{eqnarray}
\nin Equation~\ref{rdot} is invariant under the symmetry transformations
\begin{eqnarray}
\phi\rightarrow-\phi,\;\;\; \dot{\phi}\rightarrow-\dot{\phi}, \;\;\; \cl\rightarrow-\cl,\;\;\; a\rightarrow -a,\;\;\; b\rightarrow -b. \label{sym2}
\end{eqnarray}
As in the previous section, we will keep $\cl>0$ without any loss of generality. When $b>0$ ($b<0$), the magnetic force is radially out (in). Likewise, $q>0$ ($q<0$) corresponds to Coulomb repulsion (attraction). Therefore, there {\it eight} different modes of radial motion, in general. 

The weak field approximation breaks down when the electric charge or magnetic field creates curvatures comparable to that made by the black hole's mass near the event horizon. This happens when
\begin{equation}
B^2\sim M^{-2} \text{ or } Q^2\sim M^2.
\end{equation}

\nin In conventional units, the weak field approximation fails when
\begin{equation}
Q\sim \frac{G^{1/2}M}{k^{1/2}} \sim 10^{20} \frac{M}{M_\odot}\; \text{coulomb},
\end{equation}
\nin or
\begin{equation}
B\sim \frac{k^{1/2}c^3}{G^{3/2}M} \sim 10^{19} \frac{M_\odot}{M}\; \text{gauss},
\end{equation}
where $M_\odot$ is the solar mass.
\nin The typical magnetic field strength near a black hole's horizon has been estimated to be $\sim10^8$ gauss ($10^{-15} \:\text{meter}^{-1}$) for stellar mass black holes and $\sim10^4$ gauss ($10^{-19} \:\text{meter}^{-1}$) for supermassive black holes~\citep{Ruth,sil1,sil2,sil3,sil4}. According to \cite{Zaj3,Zaj2}, the charge of Sgr~A* is estimated to be in the range $10^8$--$10^{15}$ coulomb ($10^{-9}$--$10^{-2}$ meter). These estimates validate ignoring corrections to the metric due to the presence of the electromagnetic fields. 

\nin In spite of the fact that the electromagnetic fields are geometrically insignificant, their effects on the dynamics of charged particles can be significant since $e/m=2.04\times 10^{21}\; (1.11\times 10^{18})$ for electrons (protons). For electrons and protons near a black hole with $Q=10^8\; \text{coulomb}$ and $B=10^4\; \text{gauss}$, for example, 
\begin{equation}
q_{\text{e}}\sim 10^{12}\;\text{meter}, \;\;\; q_{\text{p}}\sim 10^{9}\; \text{meter},
\end{equation}
and
\begin{equation}
b_{\text{e}}\sim 10^{3}\;\text{meter}^{-1}, \;\;\; b_{\text{p}}\sim 10^{-1}\; \text{meter}^{-1}.
\end{equation}
The subscripts "e" and "p" refer to electrons and protons, respectively.

\nin Let us define the positive semi-definite function ${\cal R}(r)$ to be the right hand side of Eq.~\ref{rdot}:
\begin{eqnarray}
{\cal R}(r)&=&\varepsilon^2\frac{\Gamma_0}{r}+4aM\varepsilon\lambda-r\left[(1-\frac{2M}{r})(r^2+\lambda^2)+a^2\right].
\end{eqnarray}
\nin To determine the dynamical parameters ($\ce(r)$ and $\cl(r)$) of a circular orbit of radius $r$, we require that
\begin{eqnarray}
{\cal R}(r)=0,  \\
{\cal R}'(r)=0.
\end{eqnarray}
\nin A circular orbit is the ISCO if
\begin{eqnarray}
{\cal R}''(r_\text{o})=0.
\end{eqnarray}
\nin These three conditions become, respectively,
\begin{eqnarray}
c_5r^5+c_4r^4+c_3r^3+c_2r^2+c_1r+c_0=0, \label{con1} \\
5c_5r^4+4c_4r^3+3c_3r^2+2c_2r+c_1=0, \label{con2} \\
10c_5r_\text{o}^3+6c_4r_\text{o}^2+3c_3r_\text{o}+c_2=0. \label{con3}
\end{eqnarray}
\nin where
\begin{eqnarray}
c_0 &=& 2(3a^2bM+a\ce M-aq-\cl M)(a^2b+a\ce-\cl), \\
c_1 &=& a^2[2b(2bM^2+\cl)+\ce^2-1]-a^4b^2-4abqM-\cl^2+q^2, \\
c_2 &=& M[8ab(ab+\ce)-4b\cl+2]-2q(ab+\ce), \\
c_3 &=& 2b(\cl-a^2b)+\ce^2-1, \\
c_4 &=& 2b^2M, \\
c_5 &=& -b^2.  
\end{eqnarray}

\section{Circular Orbits around a Charged Kerr Black Hole}\label{s4}

\begin{figure*}[ht]
    \begin{center}
    \ba\non
    \hspace{.1cm}\includegraphics[width=0.32\textwidth]{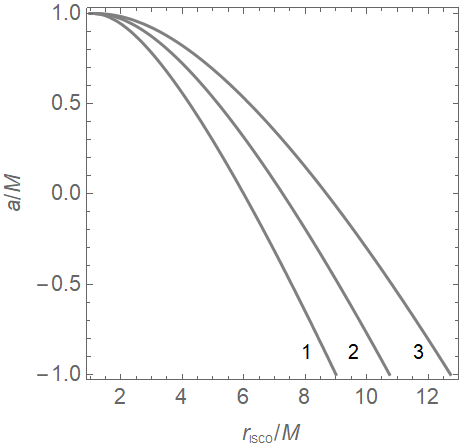}
    \hspace{.1cm}\includegraphics[width=0.32\textwidth]{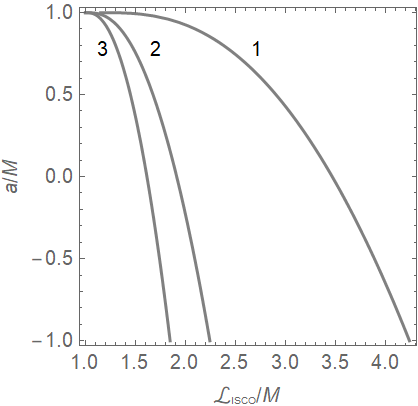}
    \hspace{.1cm}\includegraphics[width=0.32\textwidth]{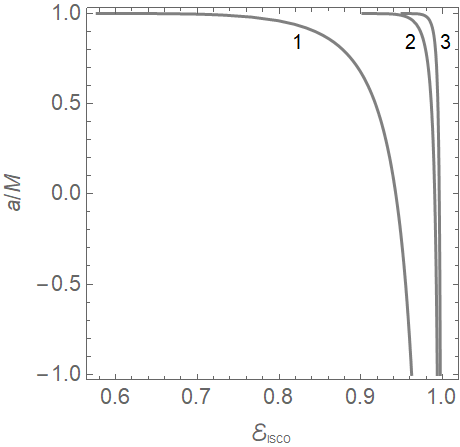}
    \ea
    \caption{The radius, specific angular momentum and specific energy of the ISCO vs $a$ for (1) $q=0$, (2) $q=0.8M$ and (3) $q=0.9M$.}\label{f:iscopq}
    \end{center}
\end{figure*}

\begin{figure*}[ht]
    \begin{center}
    \ba\non
    \hspace{.1cm}\includegraphics[width=0.32\textwidth]{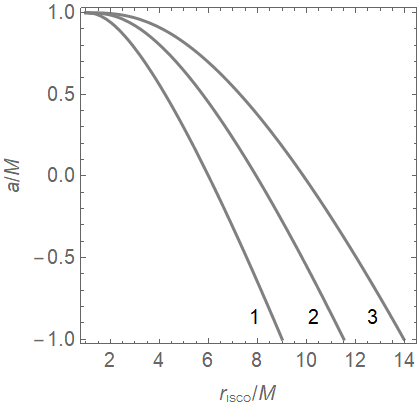}
    \hspace{.1cm}\includegraphics[width=0.32\textwidth]{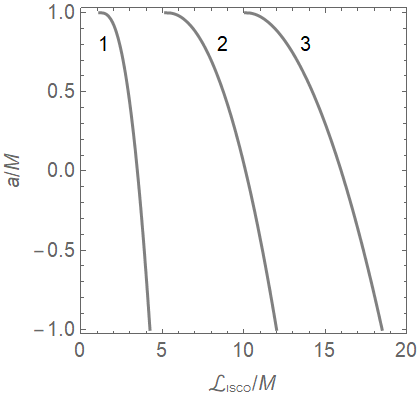}
    \hspace{.1cm}\includegraphics[width=0.32\textwidth]{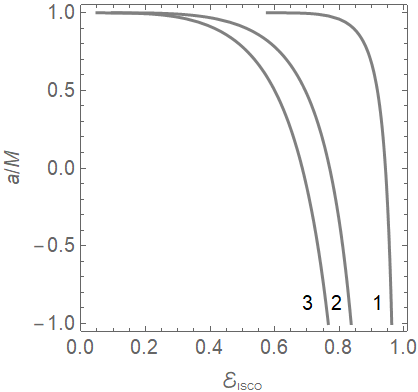}
    \ea\non
    \caption{The radius, specific angular momentum and specific energy of the ISCO vs $a$ for (1) $q=0$, (2) $q=-5.0M$ and (3) $q=-10M$.}\label{f:isconq}
    \end{center}
\end{figure*}

\nin Now, we study the circular orbits and the ISCO of a charged particle orbiting a charged Kerr black hole. We set $b=0$ in Eqs.~\ref{con1}--\ref{con3}. It is more practical to solve the resulting equations numerically. Figures~\ref{f:iscopq} and~\ref{f:isconq} show how $r_\is$, $\cl_\is$ and $\ce_\is$ vary with $a$ for selected, positive and negative values of $q$, respectively.
In both figures, $r_\is$ increases as $|q|$ increases. The increase gets steeper as $q$ approaches $M$. For $a<M$, $r_\is$ exists when $q<M$ only. When $q=M$, $r_\is$ exists only when $a=M$. In that case, $r_\is$ can take any value in the range $[M,\infty)$ with $\ce=1$ and $\cl=M$ for all values of $r_\is$. The particle is marginally stable and marginally bound at all radii. 

\nin When $q$ value is very close to $M$ ($M-q\ll M$), we can write approximate expressions for $r_\is$, $\ce_\is$ and $\cl_\is$ as
\begin{eqnarray}
r_\is &\approx & \sqrt{\frac{3(M-a)}{M-q}}M, \\
\ce_\is &\approx & 1-\left[\frac{M-q}{27(M-a)}\right]^{3/2}, \\
\cl_\is &\approx & \sqrt{3(M-a)(M-q)}+M.
\end{eqnarray}
\nin When $q$ is negative and very large in magnitude ($q\ll -M$), we can write approximate expressions for $r_\is$, $\ce_\is$ and $\cl_\is$ as
\begin{eqnarray}
r_\is &\approx & \sqrt[3]{2(M-a)q^2}, \\
\ce_\is &\approx & -3\sqrt[3]{(M-a)/4q}, \label{Eng}\\
\cl_\is &\approx & -q.
\end{eqnarray}

\nin Eq.~\ref{Eng} reveals that the efficiency of energy liberation of a charged particle ending at the ISCO can be close to $100\%$ of the particle's rest energy. The results of this section, and the relationship between $r_{\text{ISCO}}$ and $q$ in particular, demonstrates that even a trace charge on a black hole can have profound astrophysical implications.

\section{Circular Orbits around a Magnetized Kerr Black Hole}\label{s5}

\begin{figure*}[ht]
    \begin{center}
    \ba\non
    \hspace{.1cm}\includegraphics[width=0.32\textwidth]{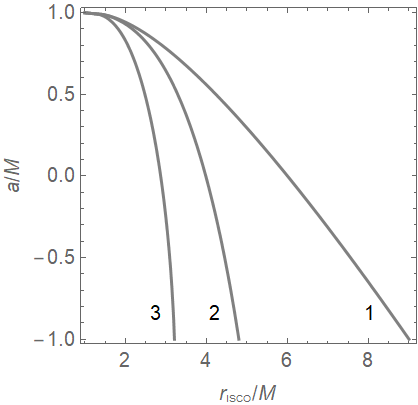}
    \hspace{.1cm}\includegraphics[width=0.32\textwidth]{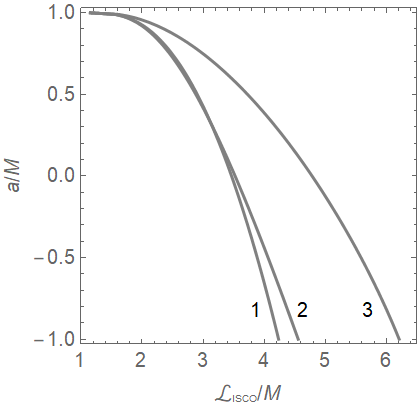}
    \hspace{.1cm}\includegraphics[width=0.32\textwidth]{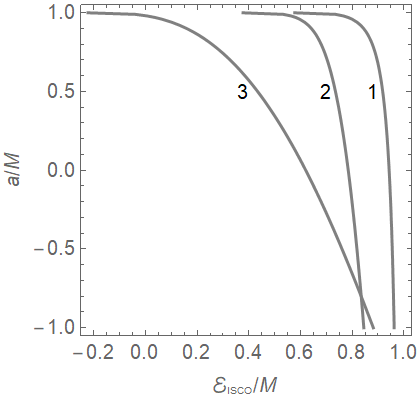}
    \ea
    \caption{The radius, specific angular momentum and specific energy of the ISCO vs $a$ for (1) $b=0$, (2) $b=0.1M^{-1}$ and (3) $b=0.4M^{-1}$.}\label{f:iscopb}
    \end{center}
\end{figure*}

\begin{figure*}[ht]
    \begin{center}
    \ba\non
    \hspace{.1cm}\includegraphics[width=0.32\textwidth]{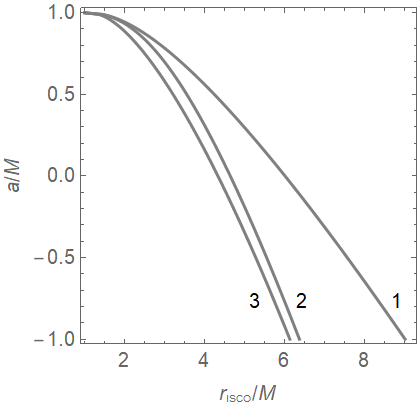}
    \hspace{.1cm}\includegraphics[width=0.32\textwidth]{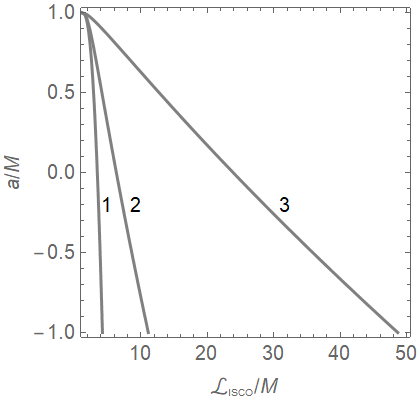}
    \hspace{.1cm}\includegraphics[width=0.32\textwidth]{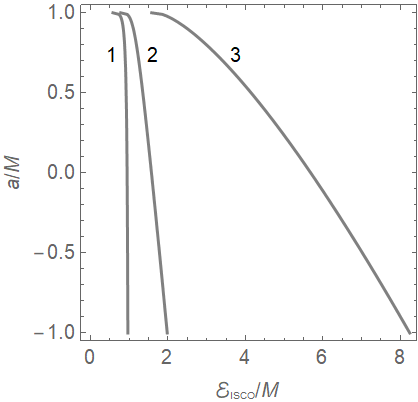}
    \ea
    \caption{The radius, specific angular momentum and specific energy of the ISCO vs $a$ for (1) $b=0$, (2) $b=-0.1M^{-1}$ and (3) $b=-0.5M^{-1}$.}\label{f:isconb}
    \end{center}
\end{figure*}

\nin In this section, we review the circular orbits and the ISCO of a charged particle near a magnetized Kerr black hole. We set $q=0$ in Eqs.~\ref{con1}--\ref{con3} and solve the resulting equations numerically.
\nin Figures~\ref{f:iscopb} and~\ref{f:isconb} show how $r_\is$, $\ce_\is$ and $\cl_\is$ vary with $a$ for selected positive and negative values of $b$, respectively. \non
\non
The effect of the magnetic field is always to bring the ISCO closer to the event horizon. Unfortunately, it does not seem possible to write simple asymptotic expressions for the ISCO parameters when $|b|\gg 1/M$. As $b\rightarrow \infty$, $r_\is$ approaches the event horizon. When $b<0$, $r_\is$ converges very quickly to some asymptotic value between $M$ and $(2^{4/3}+2^{2/3}+2)M\approx 6.107M$ depending of the value of $a$. Generally, $\ce_\is$ and $\cl_\is$ do not converge to asymptotic values as $|b|$ increases.

\begin{figure}[h!]
  \centering
  \includegraphics[width=0.45\textwidth]{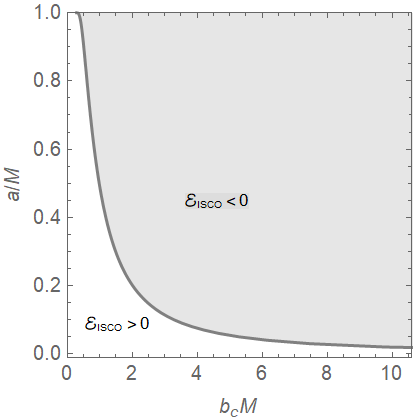}
  \caption{The dependence of the critical $b$ value after which $\ce_\is$ becomes negative $b_c$ on $a$.}\label{f:a_vs_bc}
\end{figure}

\nin The third plot in Fig~\ref{f:iscopb} demonstrates that $\ce_\is$ become negative when $b$ is greater than some critical value $b_c(a)$. When $a=M$, $b_c$ has a minimum value of $(2\sqrt{3}M)^{-1}\approx 0.289 M^{-1}$. For $a\ll M$,
\begin{equation}
b_c=\sqrt[3]{\frac{2}{3\sqrt{3}a^2M}}.
\end{equation}  
\nin Figure~\ref{f:a_vs_bc} shows how $b_c$ depends on $a$.
\nin As shown in \cite{Z}, the efficiency of energy release for a charged particle ending at the ISCO of a magnetized Kerr black hole can be several orders of magnitudes greater than its rest energy.

\section{Circular Orbits around a Charged and Magnetized Kerr Black Hole}\label{s6}

\begin{figure*}[ht]
    \begin{center}
    \ba\non
    \hspace{.1cm}\includegraphics[width=0.32\textwidth]{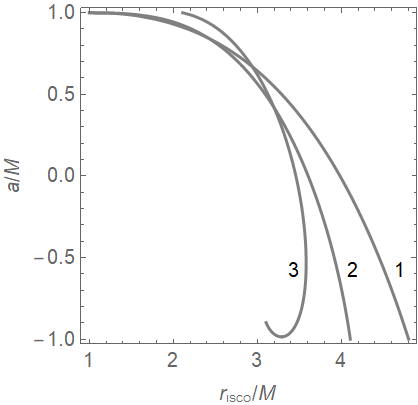}
    \hspace{.1cm}\includegraphics[width=0.32\textwidth]{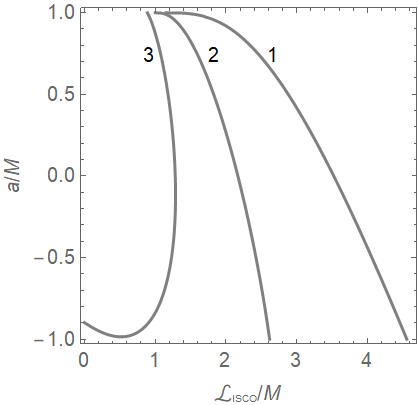}
    \hspace{.1cm}\includegraphics[width=0.32\textwidth]{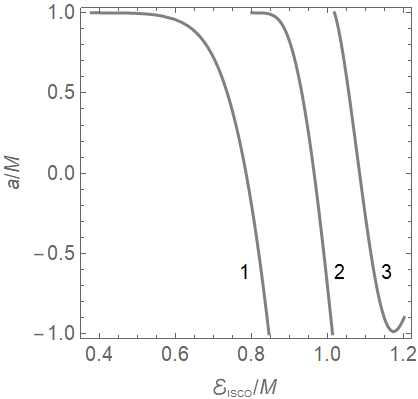}
    \ea
    \caption{The radius, specific angular momentum and specific energy of the ISCO vs $a$ for (1) $q=0$, (2) $q=1.0M$ and (3) $q=1.5M$. In all cases $b=0.1/M$.} \label{f:iscopqpb}
    \end{center}
\end{figure*}

\begin{figure*}[ht]
    \begin{center}
    \ba\non
    \hspace{.1cm}\includegraphics[width=0.32\textwidth]{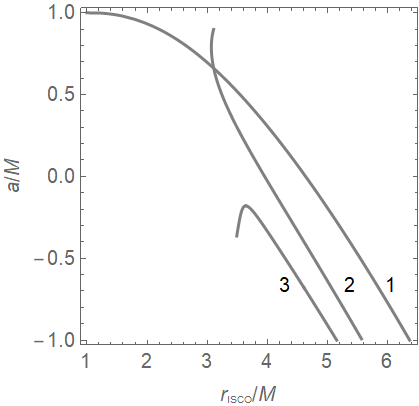}
    \hspace{.1cm}\includegraphics[width=0.32\textwidth]{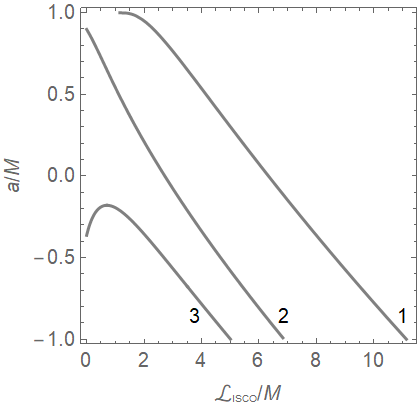}
    \hspace{.1cm}\includegraphics[width=0.32\textwidth]{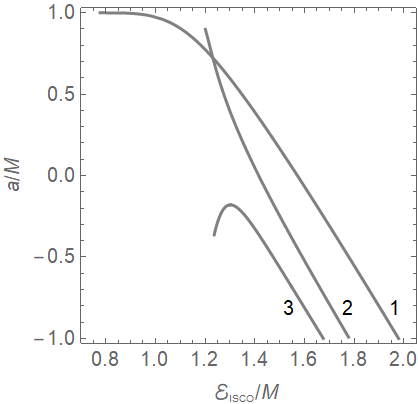}
    \ea
    \caption{The radius, specific angular momentum and specific energy of the ISCO vs $a$ for (1) $q=0$, (2) $q=1.5M$ and (3) $q=2.0M$. In all cases $b=-0.1/M$.} \label{f:iscopqnb}
    \end{center}
\end{figure*}

\begin{figure*}[ht]
    \begin{center}
    \ba\non
    \hspace{.1cm}\includegraphics[width=0.32\textwidth]{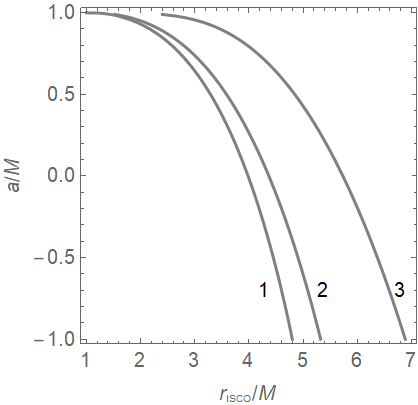}
    \hspace{.1cm}\includegraphics[width=0.32\textwidth]{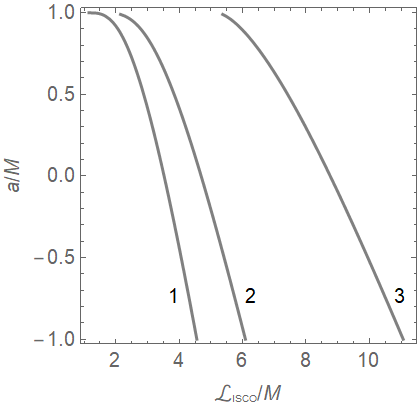}
    \hspace{.1cm}\includegraphics[width=0.32\textwidth]{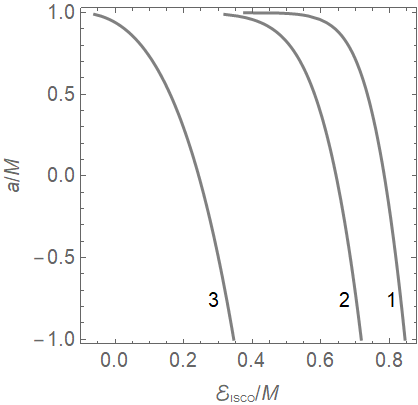}
    \ea
    \caption{The radius, specific angular momentum and specific energy of the ISCO vs $a$ for (1) $q=0$, (2) $q=-1.0M$ and (3) $q=-5.0M$. In all cases $b=0.1/M$.} \label{f:isconqpb}
    \end{center}
\end{figure*}

\begin{figure*}[ht]
    \begin{center}
    \ba\non
    \hspace{.1cm}\includegraphics[width=0.32\textwidth]{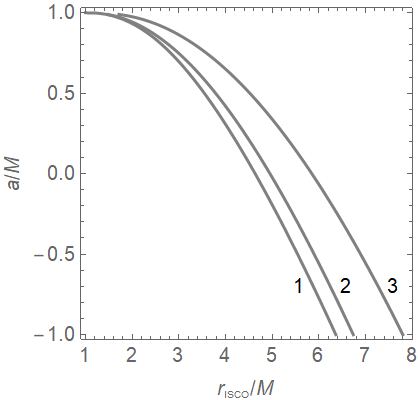}
    \hspace{.1cm}\includegraphics[width=0.32\textwidth]{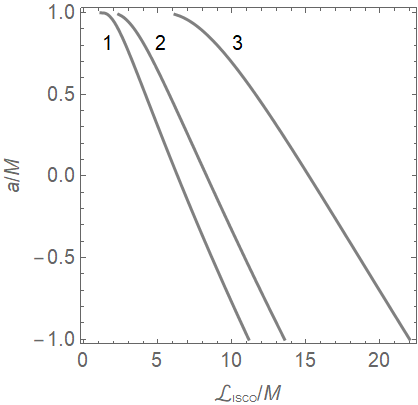}
    \hspace{.1cm}\includegraphics[width=0.32\textwidth]{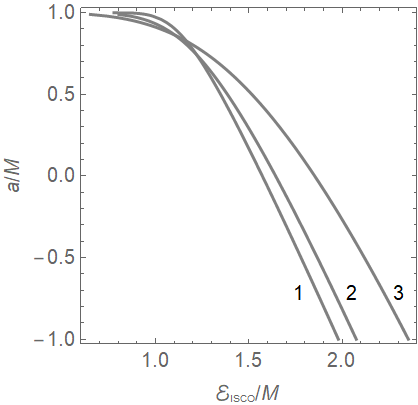}
    \ea
    \caption{The radius, specific angular momentum and specific energy of the ISCO vs $a$ for (1) $q=0$, (2) $q=-1.0M$ and (3) $q=-5.0M$. In all cases $b=-0.1/M$.} \label{f:isconqnb}
    \end{center}
\end{figure*}

\nin It is now the time to study the problem at its full complexity. In order to visualize the behavior of the ISCO parameters when both $q$ and $b$ are nonzero, we will produce plots similar to those in Figs.~\ref{f:iscopq}--\ref{f:isconb} for selected, representative values of $q$ and $b$. Figures~\ref{f:iscopqpb} and~\ref{f:iscopqnb} show the effect of turning on the magnetic field on $r_\is$, $\cl_\is$ and $\ce_\is$ vs. $a$ curves for $q>0$. Figures~\ref{f:isconqpb} and~\ref{f:isconqnb} are for the case when $q<0$.
\non
\non
\non
\non

The magnetic field has mainly three effects on $r_\is$: (1) It brings $r_\is$ closer to the black hole. In all cases when $b\neq 0$, $r_\is$ is finite. (2) It makes $q_\text{max}>M$, where $q_{\text{max}}$ is the maximum possible value of $q$ for which $r_\is$ exists. (3) It can create two concurrent ISCOs, $r_\text{in}$ and $r_\text{out}$, when $q>0$. These three effects become more evident as $|b|$ increases. The existence of two ISCOs implies the existence of a gap with no stable circular orbits or a region with two possible circular orbits. 


\nin The inner and outer ISCOs $r_\text{in}$ and $r_\text{out}$ define the inner boundaries of two distinct bands of circular orbits. The inner band ends at $r_{\text o}|_{\cl_0=0}$, the value of $r_{\text o}$ at which $\cl_{\text o}=0$. The outer band is limitless. When $r_\text{out}>r_{\text o}|_{\cl_0=0}$, the two bands do not overlap and there exists a forbidden zone where stable circular orbits cannon exist. When $r_\text{out}<r_{\text o}|_{\cl_0=0}$, the bands overlap and there exists a zone where particles in stable circular orbits of the same radius $r_{\text o}$ can have two different possibilities for angular momentum and energy.

\begin{figure}[h!]
  \centering
  \includegraphics[width=0.45\textwidth]{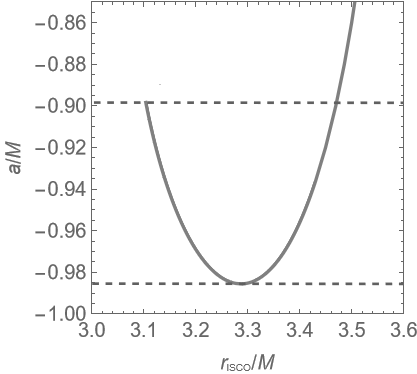}
  \caption{The double-valued part of curve 3 in the first plot of Fig.~\ref{f:iscopqpb}.}\label{f:zoom}
\end{figure}

\begin{figure}[h!]
  \centering
  \includegraphics[width=0.45\textwidth]{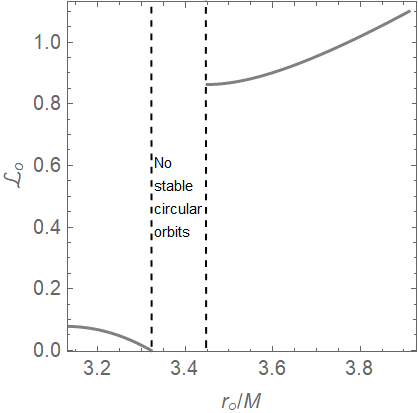}
  \caption{The dependence of $r_{\text o}$ on $\cl_{\text o}$ for $a=-0.92M$, $b=0.1/M$ and $q=1.5M$. The inner and outer bands of stable circular orbits and the no-stable-circular-orbits gap are shown.}\label{f:gap}
\end{figure}

\begin{figure}[h!]
  \centering
  \includegraphics[width=0.45\textwidth]{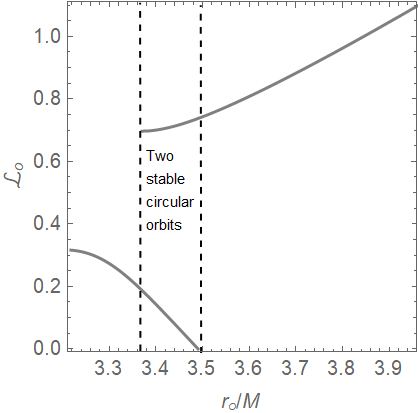}
  \caption{The dependence of $r_{\text o}$ on $\cl_{\text o}$ for $a=-0.97M$, $b=0.1/M$ and $q=1.5M$. Two possible stable circular orbits are possible in the overlap zone.}\label{f:two}
\end{figure}

\begin{figure}[h!]
  \centering
  \includegraphics[width=0.45\textwidth]{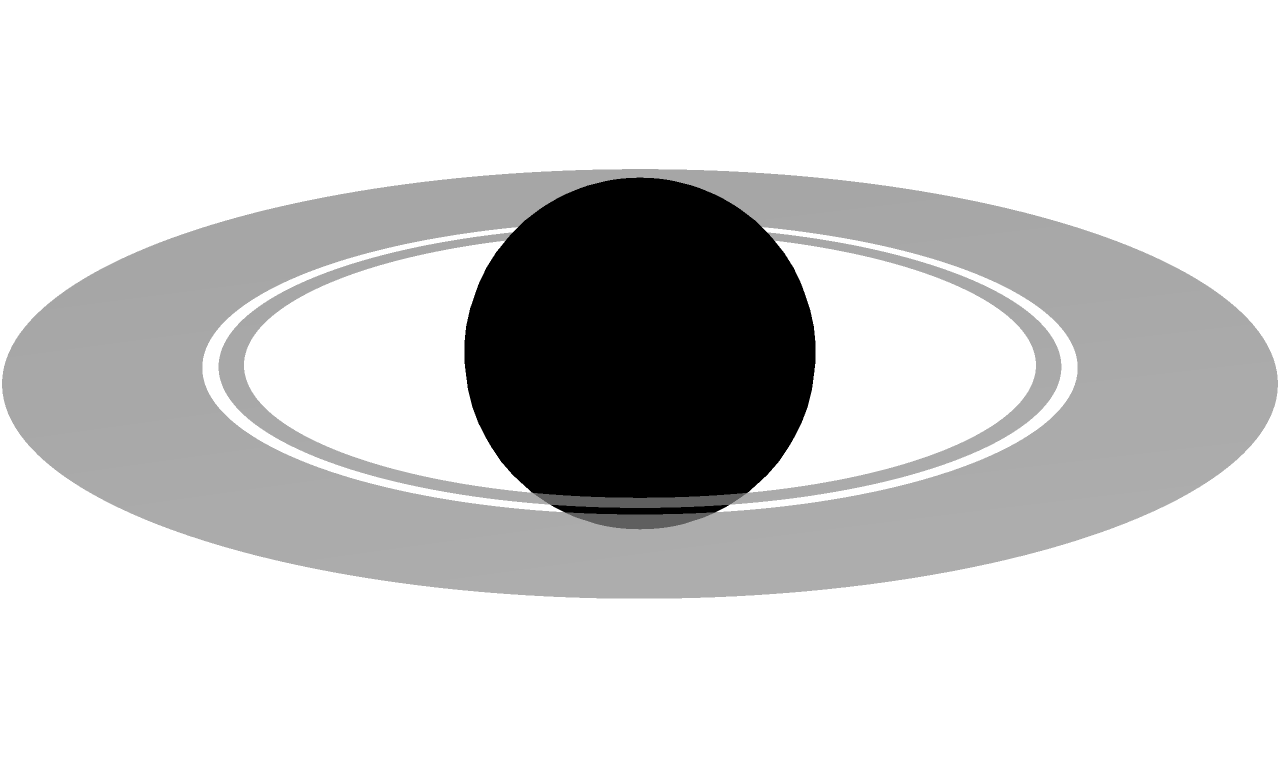}
  \caption{The black hole (black) and the two bands of circular orbits corresponding to figure~\ref{f:gap}. The gap appears clearly between the two bands. The outer band is limited to $r_{\text o}=5M$ for clarity.}\label{f:gapdisk}
\end{figure}

\begin{figure}[h!]
  \centering
  \includegraphics[width=0.45\textwidth]{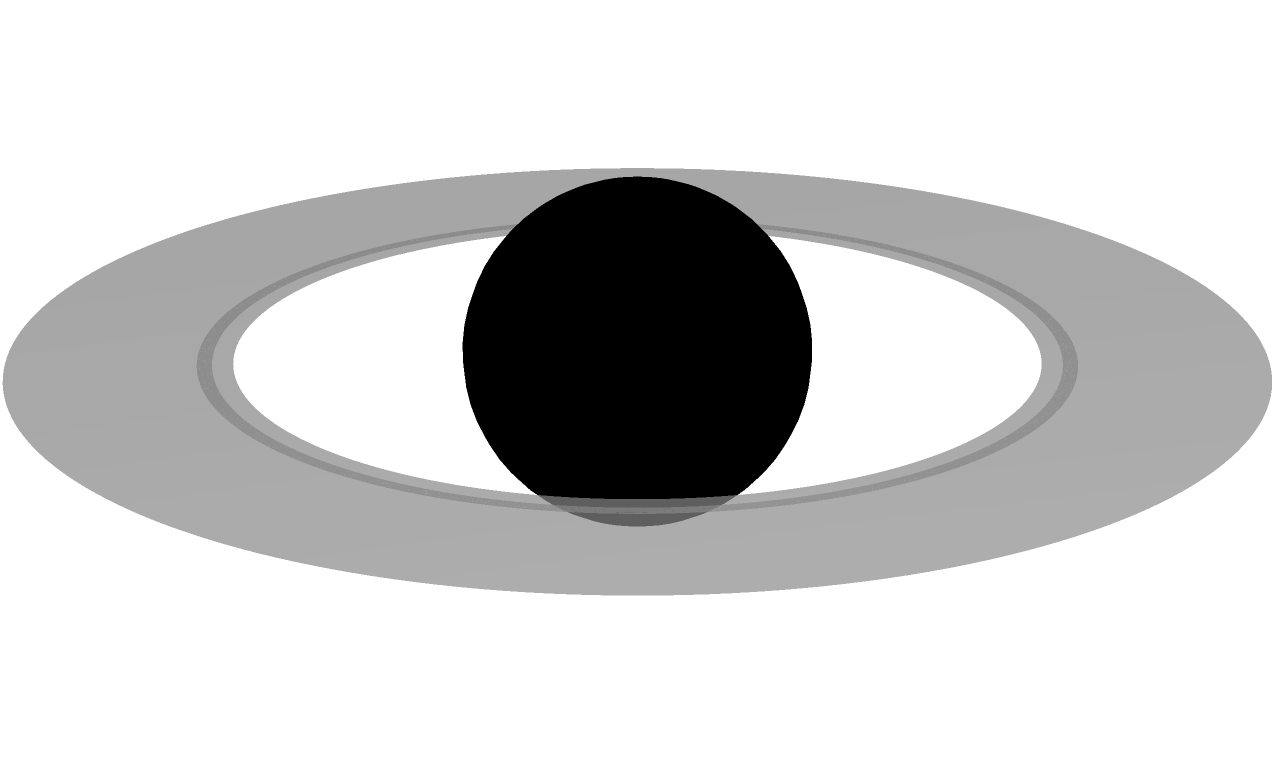}
  \caption{The black hole (black) and the two bands of circular orbits (light grey)corresponding to figure~\ref{f:two}. The double-orbit region is the intersection area of the two bands (dark grey). The outer band is limited to $r_{\text o}=5M$ for clarity.}\label{f:doubledisk}
\end{figure}

\nin For the sake of demonstration, let us focus in on the double-valued part of curve 3 in the first plot of Fig.~\ref{f:iscopqpb}. Figure~\ref{f:zoom} is a magnification of this part of the curve. For $a=-0.92M$, for example, $r_\text{in}=3.123M$, $r_\text{out}=3.449M$ and $r_{\text o}|_{\cl_0=0}=3.322M$. The dependence of $r_{\text o}$ on $\cl$ is shown in Fig.~\ref{f:gap} where the two bands and the gap appear clearly. For $a=-0.97M$, for example, $r_\text{in}=3.203M$, $r_\text{out}=3.370M$ and $r_{\text o}|_{\cl_0=0}=3.490M$. The inner and outer bands overlap in this case, as shown in Fig.~\ref{f:two}. Figures~\ref{f:gapdisk} and~\ref{f:doubledisk} are 3D plots of the circular orbits bands corresponding to Figs.~\ref{f:gap} and~\ref{f:two}, respectively. 

\nin Astrophysical black hole accretion disks have generally rich structures. However, the charged particles in low density accretion disks have relatively long free path and may be considered collisionless. The motion of the charges particles is therefore essentially Keplarian. This is the case with Sgr A* and, likely, M87, as well as many other systems \citep{Zaj3}.   

\nin Another realistic environment where collisionless charged particles can exist is dust particles in black hole systems. They may result from the tidal disruption of asteroids and planets as well. Asteroid disruption was discussed as a potential explanation for the observed flares near Sgr A* \citep{ZNM}. The irradiation of dust particle renders the dust particles positively charged via photoionization \citep{EHK}. 

\nin There is an even more realistic environment where our set-up may apply. Significant charge can be induced in the surrounding magnetosphere of accreting black holes \citep{Ruf}. This result was applied to analyse the dynamics of orbiting Galactic center flare components where the interaction of their potential charge with the ambient magnetic field was demonstrated \citep{TZE}. It may be possible to detect the potential gap in circular orbits by observing the dynamics of large-enough number of flares. The gap would then show up on, for example, a period-radius plot of the 
flares similar to that in \cite{KZK}.

\nin Black hole accretion disks with a gap were discussed in the context of supermassive black hole-black hole binaries where the less massive black hole has engraved an annular gap in the circumbinary disk~\citep{GM,RVP}. It was argued that the existence of a gap in the disk results in a clear signature in the spectral energy distribution of the disk. Even though we do not expect real accretion disks to be as simple as those in Figs.~\ref{f:gapdisk} and~\ref{f:doubledisk}, we speculate that the existence of a gap in the stable circular orbits due to the electromagnetic fields discussed in this paper will leave a clear signature in the spectral energy distribution too. We also think that the corresponding black hole shadow image will look unique. The same argument applies, maybe to a lesser extent, to accretion disks with a region of double orbit.

\begin{figure}[h!]
  \centering
  \includegraphics[width=0.45\textwidth]{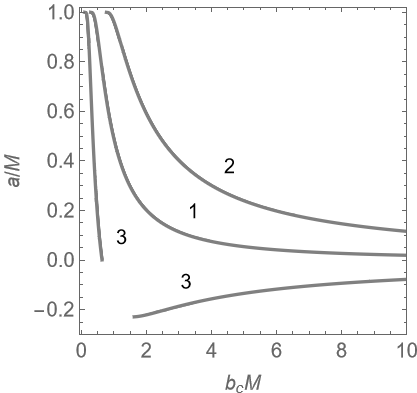}
  \caption{The dependence of the critical $b$ value after which $\ce_\is$ becomes negative $b_c$ on $a$ for (1) $q=0$, (2) $q=2M$ and (3) $q=-2M$.}\label{f:a_vs_bc_q}
\end{figure}

\nin We have seen in Sec.~\ref{s5} that negative energy stable circular orbits can exits when $b\neq 0$ and $M\geq a>0$ when $q=0$. To understand the effect of the black hole's electric charge on these orbits we reproduce the $b_c$ curve shown in Fig.~\ref{f:a_vs_bc} for selected values of $q$. The resulting $b_c$ curves are shown in Fig.~\ref{f:a_vs_bc_q}.
\nin Generally, when $q>0$ the $b_c$ is shifted up-right, which means that $b_c$ value for a certain $a$ value gets greater as $q$ gets greater. When $q<0$, two effects occur: (1) The value of $b_c$ at $a=0$ becomes finite. (2) $b_c$ exists for $a<0$. When $q$ becomes large enough in magnitude, $b_c$ exists for all values of $a$, even $a=-M$.

\section{Summary}\label{sum}

We have studied the ISCOs of charged particles near a weakly charged and magnetized Kerr black hole. Although the black hole's magnetic field may be weak and its charge may be just a trace charge, their effects on charged particles can be dominant.

\nin The effect of the black hole's charge alone is to push the ISCO beyond the neutral particle's ISCO, regardless of whether the Coulomb force is repulsive or attractive. This effect is weakened as $a$ approaches $M$. When the Coulomb force is repulsive, the ISCO does not exist beyond some critical ratio of the Coulomb force to the 'gravitational force', unless if $a=M$ and the particle is co-rotating with the black hole. In that case, the particle is marginally stable at all radii. Generally, the binding energy of a charged particle at the ISCO can be as much as the particle's rest energy. 

The effect of the magnetic field alone is to bring the ISCO closer than the neutral particle's ISCO in all cases. When the magnetic force is radially out, the particle's ISCO can approach the event horizon where the particle's binding energy can be greater that its rest energy by several orders of magnitudes. 

The problem becomes much richer when the black hole is both charged and magnetized. The charge and magnetic field have competitive effects on the ISCO's radius. The critical ratio of the Coulomb force to the 'gravitational force' beyond which the ISCO does not exit becomes greater as the magnetic field becomes stronger. 

We have seen that a tiny black hole electric charge or magnetic field can change the ISCO radius-spin angular momentum relationship drastically. Consequently, the neutral particle's ISCO radius-spin angular momentum curve may not be reliable to measure the spin angular momentum of astrophysical black holes. 

The most interesting result is the possibility of the existence of two bands of charged particles' circular orbits. The two bands may be intersecting, yielding a region where two stable circular orbits can occur. The two band may also be disjointed, separated by a region of no stable circular orbits. We think that these effects can lead to observable structures in astrophysical accretion disks.

The problem may be richer and more astrophysically interesting when more realistic electromagnetic fields are considered. The study of the exact effects of the existence of a gap and double orbits on the structure of slim accretion disks, their spectral energy distribution and their shadow images is another intriguing area to explore.

\section*{Acknowledgment}

The author gratefully acknowledges the deanship of academic research at King Fahd University of Petroleum and Minerals for financially supporting this work under project code SR181025-2. 

\bibliographystyle{aasjournal}
\bibliography{refs}{}

\end{document}